\documentclass[letterpaper,twoside,10pt]{article}
\usepackage{amssymb}
\usepackage{amsmath}
\usepackage{latexsym}
\usepackage{epsfig}
\usepackage{fancyheadings}
\usepackage{pstricks,pst-node}
\usepackage{rotating,graphicx}

\newlength{\graphscale}
\newlength{\graphunit}
\newcommand{\slg}[1]{\setlength{\graphscale}{#1\graphunit}}
\newlength{\enviropost}
\newcommand{\be}{\begin{equation}}
\newcommand{\ee}{\end{equation}}
\newcommand{\ble}[1]{\begin{equation} \label{#1}}
\newcommand{\bae}{\begin{eqnarray}}
\newcommand{\eae}{\end{eqnarray}}
\newcommand{\fle}[2]%
{\vspace{1.5ex}
\be
\label{#1}
\mbox{%
\setlength{\fboxsep}{3ex}%
\framebox{$\dss #2 $}}
\ee} 
\newcommand{\flec}[2]%
{\vspace{1.5ex}
\be
\label{#1}
\mbox{%
\setlength{\fboxsep}{3ex}%
\framebox{$\dss #2 $}}
\, \, \,  ,
\ee} 
\newcommand{\flep}[2]%
{\vspace{1.5ex}
\be
\label{#1}
\mbox{%
\setlength{\fboxsep}{3ex}%
\framebox{$\dss #2 $}}
\, \, \, .
\ee} 

\newcommand{\nn}{\nonumber}
\newcommand{\ff}{\nn \\}
\newcommand{\fe}{& = &}

\newtheorem{prop}{Proposition}
\newtheorem{lemma}{Lemma}
\newtheorem{state}{S$\! \!$}
\newtheorem{defin}{D$\! \!$}
\newtheorem{exatitle}{Example}
\newtheorem{problemdef}{Problem}
\newtheorem{soldef}{Solution}
{\noindent \textsc{Proof}:\ }%
{\hfill $\blacksquare$  \vspace{\enviropost} \\}
{\begin{soldef} 
\label{#2:Sol} 
#1
\marginpar[(\textbf{\pageref{#2:Pro}})]{(\textbf{\pageref{#2:Pro}})}
\end{soldef}}%
{\hfill $\Box$ \vspace{.5\enviropost} \\}
\newenvironment{example}[2]%
{\begin{exatitle} \label{#2} #1 \end{exatitle}}%
{\hfill $\Box$ \vspace{\enviropost} \\}
{\begin{exatitle} \label{#2} #1 \end{exatitle}}%
{\hfill $\Box$ \\}
{\begin{problemdef} 
\label{#2:Pro} 
#1 
\marginpar[(\textbf{\pageref{#2:Sol}})]{(\textbf{\pageref{#2:Sol}})}
\end{problemdef}}%
{\hfill  \vspace{.5\enviropost} \\}
{\noindent {\bf Aside:} \small}%
{\hfill \rule[-3mm]{0mm}{0mm}$\Diamond$\\}
{%
\vspace{3mm}  
\begin{center}
\begin{minipage}{.8\textwidth} 
\begin{state} 
\label{#1}%
}%
{%
\end{state}
\end{minipage}
\end{center}
\vspace{3mm}
}
{%
\vspace{3mm}  
\begin{center}
\begin{minipage}{.8\textwidth} 
\begin{defin} 
\label{#1}%
}%
{%
\end{defin}
\end{minipage}
\end{center}
\vspace{3mm}
}

\newcommand{\dss}{\displaystyle}
\newcommand{\id}{\mathop{\rm id}}

\newcommand{\ot}{\otimes}

\newcommand{\ip}[2]{\left\langle #1, #2\right\rangle}

\newcommand{\eg}{\hbox{\em e.g.{}}}
\newcommand{\etc}{\hbox{\em etc.{}}}
\newcommand{\ie}{\hbox{\em i.e.{}}}

\newcommand{\rhs}{\hbox{r.h.s.{}}}

\newcommand{\longpage}{\enlargethispage{\baselineskip}}

\newcommand{\calA}{\mathcal{A}}

\newcommand{\calH}{\mathcal{H}}

\newcommand{\calL}{\mathcal{L}}

\newcommand{\calO}{\mathcal{O}}

\newcommand{\papertitle}{%
Physics, Combinatorics and Hopf Algebras%
}
\newcommand{\paperauthor}{%
C.{} Chryssomalakos%
}
\begin{document}
\begin{titlepage}
\vspace*{-1cm}
\begin{flushright}
\textsf{}
\\
\mbox{}
\\
\textsf{April 4, 2004}
\\[3cm]
\end{flushright}
\renewcommand{\thefootnote}{\fnsymbol{footnote}}
\begin{LARGE}
\bfseries{\sffamily \papertitle}
\end{LARGE}

\noindent \rule{\textwidth}{.6mm}

\vspace*{1.6cm}

\noindent \begin{large}%
\textsf{\bfseries%
\paperauthor
}
\end{large}


\phantom{XX}
\begin{minipage}{.8\textwidth}
\begin{it}
\noindent Instituto de Ciencias Nucleares \\
Universidad Nacional Aut\'onoma de M\'exico\\
Apdo. Postal 70-543, 04510 M\'exico, D.F., M\'EXICO \\
\end{it}
\texttt{chryss@nuclecu.unam.mx
\phantom{X}}
\end{minipage}
\\

\vspace*{3cm}

\noindent
\textsc{\large Abstract: }
A number of problems in theoretical physics share a common
nucleus of combinatoric nature. It is argued here that
Hopf algebraic concepts and techiques can be particularly
efficient in dealing with such problems. As a first example, 
a brief review is given of the recent work of Connes, Kreimer and
collaborators on the algebraic structure of the process of
renormalization in quantum field theory. Then the concept of 
$k$-primitive elements is introduced --- these are particular
linear combinations of products of Feynman diagrams --- and it is
shown, in the context of a toy-model,  that they significantly reduce 
the computational
cost of renormalization.
 As a second example, Sorkin's proposal for a
family of generalizations of quantum mechanics, indexed by an
integer $k>2$, is reviewed (classical mechanics corresponds to
$k=1$, while
quantum mechanics to $k=2$). It is then shown that the quantum
measures of order $k$ proposed by Sorkin can also be described as
$k$-primitive elements of the Hopf algebra of functions on an
appropriate infinite dimensional abelian group.
\end{titlepage}
\setcounter{footnote}{1}
\renewcommand{\thefootnote}{\arabic{footnote}}
\setcounter{page}{2}
\noindent \rule{\textwidth}{.5mm}

\tableofcontents

\noindent \rule{\textwidth}{.5mm}
\newsavebox{\PROPA}
\newsavebox{\PROPALOOP}
\newsavebox{\PROPATLOOP}
\newsavebox{\PROPALLOOP}
\newsavebox{\PROPABLOB}
\sbox{\PROPA}{{%
\begin{pspicture}(0mm,-.5ex)(10mm,8mm)
\psset{xunit=1cm,yunit=1cm}
\psline[linewidth=.2mm]{-}%
(0,0)(1,0)
\end{pspicture}}%
}
\newcommand{\propa}[1][0]{%
\raisebox{#1\totalheight}%
  {\resizebox{!}{\graphscale}{\usebox{\PROPA}}}}
\sbox{\PROPALOOP}{{%
\begin{pspicture}(0mm,-.5ex)(10mm,8mm)
\psset{xunit=1cm,yunit=1cm}
\psline[linewidth=.2mm]{-}%
(0,0)(1,0)
\pscurve[linewidth=.2mm]{-}%
(.25,0)(.5,.25)(.75,0)
\end{pspicture}}%
}
\newcommand{\propaloop}[1][0]{%
\raisebox{#1\totalheight}%
  {\resizebox{!}{\graphscale}{\usebox{\PROPALOOP}}}}
\sbox{\PROPATLOOP}{{%
\begin{pspicture}(0mm,-.5ex)(17.5mm,8mm)
\psset{xunit=1cm,yunit=1cm}
\psline[linewidth=.2mm]{-}%
(0,0)(1.75,0)
\pscurve[linewidth=.2mm]{-}%
(.25,0)(.5,.25)(.75,0)
\pscurve[linewidth=.2mm]{-}%
(1,0)(1.25,.25)(1.5,0)
\end{pspicture}}%
}
\newcommand{\propatloop}[1][0]{%
\raisebox{#1\totalheight}%
  {\resizebox{!}{\graphscale}{\usebox{\PROPATLOOP}}}}
\sbox{\PROPALLOOP}{{%
\begin{pspicture}(0mm,-.5ex)(14mm,8mm)
\psset{xunit=1cm,yunit=1cm}
\psline[linewidth=.2mm]{-}%
(0,0)(1.4,0)
\pscurve[linewidth=.2mm]{-}%
(.25,0)(.7,.45)(1.15,0)
\pscurve[linewidth=.2mm]{-}%
(.45,0)(.7,.25)(.95,0)
\end{pspicture}}%
}
\newcommand{\propalloop}[1][0]{%
\raisebox{#1\totalheight}%
  {\resizebox{!}{\graphscale}{\usebox{\PROPALLOOP}}}}
\sbox{\PROPABLOB}{{%
\begin{pspicture}(0mm,-.5ex)(10mm,8mm)
\psset{xunit=1cm,yunit=1cm}
\psline[linewidth=.2mm]{-}%
(0,0)(.2,0)
\psline[linewidth=.2mm]{-}%
(.8,0)(1,0)
\pscircle[linewidth=.2mm,fillstyle=vlines,%
hatchangle=45,hatchwidth=.4pt,hatchsep=2pt,fillcolor=black](.5,0){.3}
\end{pspicture}}%
}
\newcommand{\propablob}[1][0]{%
\raisebox{#1\totalheight}%
  {\resizebox{!}{\graphscale}{\usebox{\PROPABLOB}}}}
\newlength{\treescale}
\newlength{\treeunit}
\setlength{\treeunit}{1ex}
\newcommand{\slt}[1]{\setlength{\treescale}{#1\treeunit}}
\newcommand{\slte}{\setlength{\treescale}{1.8\treeunit}}
\newcommand{\sltt}{\setlength{\treescale}{1.5\treeunit}}
\newcommand{\slts}{\setlength{\treescale}{1.4\treeunit}}
\newcommand{\sltss}{\setlength{\treescale}{1.2\treeunit}}
\newsavebox{\Treeoo}
\newsavebox{\Treeto}
\newsavebox{\Treetho}
\newsavebox{\Treetht}
\newsavebox{\Treefo}
\newsavebox{\Treeft}
\newsavebox{\Treefth}
\newsavebox{\Treeff}
\newsavebox{\Treefio}
\newsavebox{\Treefit}
\newsavebox{\Treefith}
\newsavebox{\Treefif}
\newsavebox{\Treefifi}
\newsavebox{\Treefis}
\newsavebox{\Treefise}
\newsavebox{\Treefie}
\newsavebox{\Treefin}
\sbox{\Treeoo}{{%
\begin{pspicture}(2mm,-1mm)(8mm,10mm)
\psset{xunit=1cm,yunit=1cm}
\psdots[dotscale=3](.5,0)
\end{pspicture}}%
}
\newcommand{\Too}[1][0]{%
\raisebox{#1\totalheight}%
  {\resizebox{!}{\treescale}{\usebox{\Treeoo}}}}
\sbox{\Treeto}{{%
\begin{pspicture}(2mm,-1mm)(8mm,10mm)
\psset{xunit=1cm,yunit=1cm}
\psline[linewidth=.3mm]{*-*}%
(.5,1)(.5,0)
\psdots[dotscale=3](.5,1)(.5,0)
\end{pspicture}}%
}
\newcommand{\Tto}[1][0]{%
\raisebox{#1\totalheight}%
  {\resizebox{!}{\treescale}{\usebox{\Treeto}}}}
\sbox{\Treetho}{{%
\begin{pspicture}(2mm,-1mm)(8mm,2cm)
\psset{xunit=1cm,yunit=1cm}
\psline[linewidth=.3mm]{*-*}%
(.5,2)(.5,0)
\psdots[dotscale=3](.5,2)(.5,1)(.5,0)
\end{pspicture}}%
}
\newcommand{\Ttho}[1][0]{%
\raisebox{#1\totalheight}%
  {\resizebox{!}{2\treescale}{\usebox{\Treetho}}}}
\sbox{\Treetht}{{%
\begin{pspicture}(-3mm,-1mm)(13mm,10mm)
\psset{xunit=1cm,yunit=1cm}
\psline[linewidth=.3mm]{*-*}%
(.5,1)(0,0)
\psline[linewidth=.3mm]{*-*}%
(.5,1)(1,0)
\psdots[dotscale=3](.5,1)(0,0)(1,0)
\end{pspicture}}%
}
\newcommand{\Ttht}[1][0]{%
\raisebox{#1\totalheight}%
  {\resizebox{!}{\treescale}{\usebox{\Treetht}}}}
\sbox{\Treefo}{{%
\begin{pspicture}(2mm,-1mm)(8mm,30mm)
\psset{xunit=1cm,yunit=1cm}
\psline[linewidth=.3mm]{*-*}%
(.5,3)(.5,0)
\psdots[dotscale=3](.5,3)(.5,2)(.5,1)(.5,0)
\end{pspicture}}%
}
\newcommand{\Tfo}[1][0]{%
\raisebox{#1\totalheight}%
  {\resizebox{!}{3\treescale}{\usebox{\Treefo}}}}
\sbox{\Treefth}{{%
\begin{pspicture}(2mm,-1mm)(13mm,20mm)
\psset{xunit=1cm,yunit=1cm}
\psline[linewidth=.3mm]{*-*}%
(.5,2)(.5,0)
\psline[linewidth=.3mm]{*-*}%
(.5,2)(1,1)
\psdots[dotscale=3](.5,2)(.5,1)(.5,0)(1,1)
\end{pspicture}}%
}
\newcommand{\Tfth}[1][0]{%
\raisebox{#1\totalheight}%
  {\resizebox{!}{2\treescale}{\usebox{\Treefth}}}}
\sbox{\Treeft}{{%
\begin{pspicture}(-3mm,-1mm)(13mm,20mm)
\psset{xunit=1cm,yunit=1cm}
\psline[linewidth=.3mm]{*-*}%
(.5,2)(.5,1)
\psline[linewidth=.3mm]{*-*}%
(.5,1)(0,0)
\psline[linewidth=.3mm]{*-*}%
(.5,1)(1,0)
\psdots[dotscale=3](.5,2)(.5,1)(0,0)(1,0)
\end{pspicture}}%
}
\newcommand{\Tft}[1][0]{%
\raisebox{#1\totalheight}%
  {\resizebox{!}{2\treescale}{\usebox{\Treeft}}}}
\sbox{\Treeff}{{%
\begin{pspicture}(-3mm,-1mm)(13mm,10mm)
\psset{xunit=1cm,yunit=1cm}
\psline[linewidth=.3mm]{*-*}%
(.5,1)(0,0)
\psline[linewidth=.3mm]{*-*}%
(.5,1)(.5,0)
\psline[linewidth=.3mm]{*-*}%
(.5,1)(1,0)
\psdots[dotscale=3](.5,1)(0,0)(.5,0)(1,0)
\end{pspicture}}%
}
\newcommand{\Tff}[1][0]{%
\raisebox{#1\totalheight}%
  {\resizebox{!}{\treescale}{\usebox{\Treeff}}}}
\sbox{\Treefio}{{%
\begin{pspicture}(2mm,-1mm)(8mm,40mm)
\psset{xunit=1cm,yunit=1cm}
\psline[linewidth=.3mm]{*-*}%
(.5,4)(.5,0)
\psdots[dotscale=3](.5,4)(.5,3)(.5,2)(.5,1)(.5,0)
\end{pspicture}}%
}
\newcommand{\Tfio}[1][0]{%
\raisebox{#1\totalheight}%
  {\resizebox{!}{\treescale}{\usebox{\Treefio}}}}
\section{Introduction}
\label{Intro}
\slts
This paper deals with  cases where
combinatoric problems arising in physics may be efficiently 
handled with geometric means. 
Two particular examples are used to illustrate the point: the
quest for primitive elements in the Hopf algebra of
renormalization of Connes and Kreimer~\cite{Con.Kre:98,Con.Kre:00}, 
on one hand, 
and the classification
of possible generalizations of quantum mechanics, proposed by
Sorkin~\cite{Sor:97,Sor:97a}, according to
the additivity properties of the corresponding ``quantum
measure'', on the other. In both cases, an underlying 
infinite-dimensional Lie group structure
permits the geometrization of the problem, as a result of
which combinatoric operations are handled by differential
geometric machinery, both facilitating and illuminating its
solution. In the first case, the group is
the non-abelian group of renormalization schemes, 
introduced by 
Connes and Kreimer, while in the second, it is the abelian group 
of characteristic functions of subsets of the set of histories of 
a quantum particle, introduced in~\cite{Chr.Dur:03}. The 
renormalization problem is presented in
Sect.~3, the source of the results
being~\cite{Chr.Que.Ros.Ver:01}. Generalized quantum mechanics appears
in Sect.~4, the exposition following~\cite{Chr.Dur:03}. Before 
that, in Sect.~2, a Hopf algebra primer
translates some familiar Lie group concepts into a language
suitable for the applications at hand. A final section of
conclusions is there  to transmit a sense of order.   
\section{A Hopf Algebra Primer}
\label{HAP}
The amount of Hopf algebraic machinery needed in the sequel is
quite modest. In fact, it is no more than what the
concept of a Lie group supplies, appropriately dualized. Thus,
in principle, everything in this paper could be cast in
familiar Lie group language, but at the cost of significant
notational inconvenience. I opt for a quick translation of Lie
groups into Hopf algebraic terms --- the reader will find that it
is well-worth the modest initial investment.

The definition of a Lie group $G$ entails the concepts of a {\em
product} $\cdot$, a {\em unit} $e$, and an {\em inverse} $g^{-1}$, 
together with appropriate 
smoothness conditions. These are operations defined in terms
of the {\em points} of the group manifold: given two points,
one can associate to them their product, there is a special
point which is neutral with respect to multiplication \etc.
Admitting linear combinations of group elements, with real
coefficients, one obtains the {\em group algebra}
$\mathbb{R}(G)$, with the
product inherited from that of the group, and defined to
distribute over the addition, so that, \eg, $(\lambda_1
g_1+\lambda_2 g_2) \cdot g_3 = \lambda_1 (g_1 \cdot g_3) +
\lambda_2 (g_2 \cdot g_3)$, $\lambda_i \in \mathbb{R}$, $g_i
\in G$.  

Dual to the vector space generated by the points of $G$, is the 
vector space 
$\text{Fun}(G)$ of 
functions on $G$. The duality is via a bilinear inner product
$\ip{\cdot}{\cdot} \colon  \mathbb{R}(G) \otimes \text{Fun}(G)
\rightarrow \mathbb{R}$, given simply
by the evaluation of the function on the point, 
$g \otimes f \mapsto \ip{g}{f} \equiv
f(g)$. Notice that the space of functions is
endowed with an algebra structure as well, given by free
commutativity. Now, one has learned in quantum
mechanics that, given an inner product between two vector spaces and
an operator acting on one of them, one can define its adjoint, 
acting on the other. In an analogous fashion, one may dualize
the above defining operations of a Lie group to corresponding
operations on the functions on the group. To make this look
nice, everything is expressed as maps. The product in $G$
is then a bilinear map $m \colon G \otimes G \rightarrow G$, 
$m(g_1 \otimes g_2) \equiv g_1 \cdot g_2$ 
(we may define $m$ on $G
\otimes G$, rather than $G \times G$ because of bilinearity).
Similarly, the unit is formalized into a map $\eta: \,
\mathbb{R} \rightarrow \mathbb{R}(G)$, $\lambda \mapsto
\eta(\lambda) = \lambda e$, \ie, $\eta$ sends every number
to that same number times the unit of the group. 
Finally, the inverse is promoted to the map
$S \colon \mathbb{R}(G) \rightarrow \mathbb{R}(G)$, $g \mapsto
S(g) = g^{-1}$, extended by {\em linearity} to the entire
$\mathbb{R}(G)$. This last feature may sound weird for an
inverse map, and indeed it is as it implies, \eg, that
$S(g_1+g_2)=S(g_1) + S(g_2) = g_1^{-1}+g_2^{-1}$, or, even
worse, $S(\lambda g) = \lambda g^{-1}$. To avoid confusion,
$S$ is called the {\em antipode} map, with the property that
it returns the group inverse when evaluated on the group
elements themselves. 

The stage is set now for dualizing
everything. One may start with the easiest, the unit map $\eta$.
Its dual is the {\em counit} map $\epsilon \colon \text{Fun}(G)
\rightarrow \mathbb{R}$, $f \mapsto \epsilon(f) = f(e)$, \ie,
the counit of a function is a number, its value at the identity  
(notice how dualizing reverses the direction of the arrows in
the map definitions). Next comes the dual of the antipode,
also called the antipode, and also denoted by $S$, 
\ble{Sdual}
\ip{S(g)}{f}= \ip{g^{-1}}{f} \equiv \ip{g}{S(f)}
\, ,
\ee
\ie, the antipode $S(f)$ of a function $f$, is a function
that, when evaluated on a
point $g$, returns the value of $f$ on $g^{-1}$. Just like
its dual, $S$ extends by linearity to the whole of
$\text{Fun}(G)$. Finally, dual to the product map is the {\em
coproduct} map $\Delta: \, \text{Fun}(G) \rightarrow
\text{Fun}(G) \otimes \text{Fun}(G)$, $f \mapsto \Delta(f)
\equiv \sum_i f_{(1)}^i \otimes f_{(2)}^i \equiv f_{(1)}
\otimes f_{(2)}$, such that
\bae
\label{Ddef}
\ip{m(g_1 \otimes g_2)}{f} 
\fe 
\ip{g_1 \otimes g_2}{\Delta(f)}
\ff
 & \equiv &
\ip{g_1 \otimes g_2}{f_{(1)} \otimes f_{(2)}}
\ff
 & \equiv & \ip{g_1}{f_{(1)}} \ip{g_2}{f_{(2)}}
\, .
\eae
Some remarks on the notation used might be helpful. The
coproduct of a function $f$ is $\Delta(f)$, a function of {\em two} 
arguments, such that $\Delta(f)(g_1,g_2) = f(g_1 g_2)$ (I will
drop occasionally the dot from the group product). Now, the
tensor product of two functions, of a single argument each,
can be considered as
a function of two arguments, $(f \otimes f')(g_1,g_2) 
= \ip{g_1 \otimes g_2}{f \otimes f'}= f(g_1)f'(g_2)$, where the last
equation {\em defines} the inner product between $\mathbb{R}(G) \otimes
\mathbb{R}(G)$ and $\text{Fun}(G) \otimes \text{Fun}(G)$. A
general function of two arguments though, and, in particular, 
$\Delta(f)$, does not necessarily
factorize like this but involves instead a sum over such
tensor products, written above as $\sum_i f_{(1)}^i \otimes
f_{(2)}^i$ for the case of $\Delta(f)$. To further enhance confusion,
one usually drops the summation symbol and the associated
index and writes simply $\Delta(f)=f_{(1)} \otimes f_{(2)}$, a
powerful notation due to Sweedler. 

An example might be due to illustrate the above. Consider the group
$A(1)$
of affine transformations of the real line, \ie, of maps
$(a,b)$: $x \mapsto a x +b$, with $a,b$ real and $a>0$. The
group  law is given by composition, 
\ble{grlaw}
(a,b) \cdot (a',b')
\equiv 
(a,b) \circ (a',b')
=
(a a', ab'+b)
\, ,
\ee
with the unit
being the point $(1,0)$ and the inverse given by
\ble{invA1}
(a,b)^{-1}= (a^{-1}, -a^{-1}b)
\, .
\ee
Introduce now coordinate functions $f$, $h$ on the group
manifold (the right half-plane), such that
\ble{fhdef}
\ip{(a,b)}{f}= f \big( (a,b) \big) =a
\, ,
\qquad
\ip{(a,b)}{h}= h \big( (a,b) \big) =b
\, .
\ee
What is the coproduct of $f$? A glance at~(\ref{grlaw}) shows
that $\Delta(f) = f \otimes f$. Indeed,
\bae
\label{copfder}
f \left( (a,b) \cdot (a',b') \right)
& \equiv &
\ip{(a,b) \cdot (a',b')}{f}
\ff
 \fe
\ip{(a,b) \otimes (a',b')}{\Delta(f)}
\ff
 \fe
\ip{(a,b) \otimes (a',b')}{f \otimes f}
\ff
 \fe
\ip{(a,b)}{f} \ip{(a',b')}{f}
\ff
 \fe
a a'
\ff
 \fe
f \left( (a a', a b' +b) \right)
\, .
\eae
Similarly one finds $\Delta(h)=f \otimes h + h \otimes 1$,
where the unit denotes the constant unit function on $A(1)$.
The counits are
\ble{couA1}
\epsilon(f)=f \big( (1,0) \big)=1
\, ,
\qquad
\epsilon(h)=h \big( (1,0) \big)=0
\, .
\ee
Finally, from~(\ref{invA1}) one  infers the antipode,
\ble{antiA1}
S(f) = \frac{1}{f}
\, ,
\qquad
S(h) = - \frac{h}{f}
\, .
\ee
\section{Primitive Elements in the Hopf Algebra of
Renormalization}
\label{Renorm}
\subsection{The need for renormalization}
\label{nfr}
Imagine playing an underwater bowling game. Any acceleration of the
ball entails the acceleration of part of the water surrounding
it, so that the mass $m_{\text{game}}$ that you observe is the
mass $m_0$ of the ball plus a correction $\Delta m_0$, due to
its interaction with the water. Or, consider
a spherical conductor of mass $m_0$. A charge $q$ is added
to the conductor which is now subjected to acceleration. The
electrical self-interaction results in a net force opposing the
acceleration, proportional (approximately) to the
acceleration.
Again, the observed mass will contain
corrections to $m_0$ due to the interaction with the
electromagnetic field. 

Physicists studying these systems have the option to 
write their dynamical equations in terms of the
original, {\em bare} quantities ($m_0$), or the observed ones 
($m_{\text{game}}$) --- the
passage between the two constitutes a
{\em renormalization} of the theory. Using either set of
quantities makes sense physically, because the interaction, in
the above examples, can be turned off 
(playing bowling in the air, or discharging the
conductor) so that either quantity is, at least in principle, 
measurable. Notice also that in both cases the correction $\Delta m$ 
is finite,
showing that renormalization is not intrinsically related to
infinities.

Renormalization also appears in quantum field theory, but
there it is a necessity, not an option. This is due to two
reasons: first, the self-interactions of the quantum fields
cannot be turned off, so that the bare quantities,
corresponding to the free field case, are fictitious, only
measurable in thought experiments. Second,
the corrections to these fictitious quantities due to
self-interactions are often infinite. Perturbative quantum
field theory, nevertheless, treats the interactions as
perturbations to the free field case. For example, in $\phi^3$
theory, where the lagrangian density is given by
\ble{Lphi3}
\calL = \frac{1}{2} \partial^\mu \phi \partial_\mu \phi 
-\frac{1}{2} m_0^2 \phi^2 
- \frac{\lambda_0}{3!} \phi^3
\, ,
\ee
the last term is the self-interaction and its effects are
calculated perturbatively. If $\lambda_0$ were zero, the observed mass
of the corresponding particle would be $m_0$, the constant that
appears explicitly in $\calL$. But $\lambda_0$ is not zero, and
the amplitude for propagation from event $A$ to event $B$ 
involves 
a sum over all possible trajectories compatible with the
initial and final positions, including intermediate 
processes of emission
and subsequent reabsorption of virtual particles.
Graphically this looks like 
\slg{8}
\bae
\label{Gamp}
\propablob
\fe
\propa + \propaloop 
\ff
 & & {}+ \propatloop + \propalloop + \ldots
\nonumber
\eae
Each of the diagrams in the \rhs{} above stands for a
particular, usually divergent, integral, which contributes to
the mass correction. Infinities (in one-particle-irreducible diagrams) 
can be {\em primitive} (\eg, second
term in the \rhs{} above) or
{\em nested} (\eg, last term). The renormalization procedure 
consists of
two steps: first {\em regularize}, \ie, find a (totally
artificial) method to make all integrals converge, for
example, by truncating the integration domain in momentum
space down to some finite size. Second,
rewrite the Lagrangian in terms of the observable parameters
and remove the regularization. The expression of the
observable parameters in terms of the bare ones will now
involve (unobservable) infinities but the functional dependence 
of the Lagrangian on the former will not. If this can be done
consistently to all orders the theory is {\em renormalizable}.
\subsection{A toy model}
\label{atm}
Illustrating the above in the context of a realistic field
theory involves rather hideous algebra. A good deal of the
intricacies of the process though is captured by the following
toy model~\cite{Bro.Kre:99}. 
There is a single primitive divergent integral $I(c)$, given by
\ble{Idef}
I(c) = \int_0^\infty \frac{dy}{y+c}
\, ,
\ee
where $c$ will be referred to as the {\em external parameter}.
We represent this graphically by a single dot, $\slte I(c)=\Too \, $.
Nested divergences are obtained by nesting $I(c)$. The
simplest case is 
\ble{I2def}
I_2(c)=
\int_0^\infty 
\frac{dy}{y+c}
\int_0^\infty
\frac{dz}{z+y}
\, ,
\ee
where the nesting is effected by letting the  external parameter of 
the second appearance of $I(c)$ be the integration variable of
the first --- the corresponding graph is $\slte \Tto$, where
the top dot refers to the first integral in~(\ref{I2def}) and
the bottom one to the one nested inside it. When nesting
twice, two possibilities exist
\bae
\label{I312def}
I_{3_1}
\fe
\int_0^\infty
\frac{dy}{y+c}
\int_0^\infty
\frac{dz}{z+y}
\int_0^\infty
\frac{dw}{w+z}
\ff
I_{3_2}
\fe
\int_0^\infty
\frac{dy}{y+c}
\int_0^\infty
\frac{dz}{z+y}
\int_0^\infty
\frac{dw}{w+y}
\, ,
\eae
with corresponding graphs $\slte \Ttho$, $\slte \Ttht$ respectively.
Clearly, an infinite family of divergent integrals is
produced, indexed by graphs known as {\em rooted trees} ({\em
trees} because there are no closed circuits in them, {\em
rooted} because the top vertex ($=$ root) is special in that it has no
parents) --- these are the Feynman diagrams of our toy model. 

We apply now a renormalization procedure to the above
integrals. The regularization is by modifying the measure, $dy
\rightarrow y^{- \epsilon} dy$. The resulting convergent
integrals can be expanded in a Laurent series in $\epsilon$,
the poles of which are recursively removed. For $I(c)$ one gets
\bae
\label{Icren}
I(c) \rightarrow I(\epsilon; c)
\fe
\int_0^\infty \frac{y^{-\epsilon}dy}{y+c}
\ff
 \fe
\frac{1}{\epsilon} - a + \calO(\epsilon)
\, ,
\eae
where $a \equiv \ln c$. The renormalized integral is obtained
by subtracting from $I(\epsilon;c)$ its pole part and letting
$\epsilon$ go to zero,
\ble{Irendef}
I^{\text{ren}}(c) = \lim_{\epsilon \rightarrow 0} 
\left(
I(\epsilon;c) - \frac{1}{\epsilon} 
\right) = -a
\, .
\ee
For the nested integrals, things are a little more
complicated. Regularization proceeds as before but the pole
subtraction has to be done recursively, starting with the
innermost (``bottom'', in the tree representation) divergences
and working one's way up the tree towards the root. The
renormalized value of the integral is given by a sum of $2^n$
terms, where $n$ is the number of nested subdivergences, equal
to the number of vertices of the corresponding tree --- a very 
clear algorithmic description of the process 
can be found in~\cite{Bro.Kre:99}. 

As one can appreciate already from the above sketchy
presentation, the renormalization process is far from unique.
One may clearly regularize the integrals in an infinite number of
ways and even the pole subtraction procedure can involve, for
example, subtracting a finite part along with the poles. The
resulting renormalized values of the integrals depend on these
choices --- it is one of the subtleties of the subject that
the physics does not. To make this a little more plausible,
consider an infinite homogeneous linear charge density and
compute the electrostatic potential a radial distance $\rho=c$ 
away from the charge. Proceeding na\"{\i}vely, one puts the zero of the
potential at $\rho=\infty$ and integrates (minus) the electric field
from infinity to $c$ to find, in appropriate units
\ble{Vex}
V(c)= -\int_\infty^c \frac{d\rho}{\rho}=\int_0^\infty
\frac{dy}{y+c}
\, ,
\ee
our old friend! The problem is not in the physics, but in the
choice of the reference point for the potential. To deal with it, 
one may choose a regularization $V(\epsilon;c)$ (as we did 
with $I(c)$ above),
 and then
move the reference point to some finite distance $\rho_0$ by
forming the difference $V(\epsilon;\rho) -
V(\epsilon;\rho_0)$ and taking the limit $\epsilon \rightarrow
0$. In so doing, one is essentially
subtracting the pole from $V(\epsilon;\rho)$, {\em plus an
arbitrary finite amount which depends on $\rho_0$}. The
physics, of course, lies in differences in the potential and,
hence, is unaffected by changes in the subtraction procedure.
A particular choice of regularization plus pole-removal
procedure is
referred to as a {\em renormalization scheme}. We summarize: 
\begin{itemize}
\item
there are
infinitely many renormalization schemes to chose from
\item
the physics does
not depend on the choice
\end{itemize} 
\subsection{Geometrization, part I}
\label{gpI}
Consider the commutative algebra $\calH$ generated by an infinite 
family of functions $\{
\phi^T \}$, with $T$ ranging over all rooted trees.
A {\em character} on $\calH$ is a linear map $\chi: \, \calH
\rightarrow \mathbb{R}$ that respects the algebra structure,
\ble{charprop}
\chi(\phi^{T_1} \phi^{T_2})= \chi(\phi^{T_1})
\chi(\phi^{T_2})
\, .
\ee
Call $G$ the space of characters on
$\calH$. You can think of the $\phi$'s as functions on $G$,
the value of $\phi^T \in \calH$ on a particular point 
$\chi \in G$ being given by $\phi^T(\chi) \equiv \chi(\phi^T)$.
Then the character property~(\ref{charprop}) corresponds to
pointwise multiplication of the $\phi$'s, \ie, 
$(\phi^{T_1} \phi^{T_2})(\chi)=
\phi^{T_1}(\chi) \phi^{T_2}(\chi)$. Now, the renormalized 
value, in a certain renormalization scheme, of the
product of two divergent integrals is defined to be the
product of the renormalized values (in the same scheme) 
of the factors. This means that renormalization schemes are
points in $G$. The value of $\phi^T$ on a point (scheme) $g$ is the
renormalized value of the integral corresponding to the tree
$T$ in the scheme corresponding to $g$. Furthermore, the 
$\phi$'s are coordinate
functions on $G$, since, to completely specify a character,
one only needs its values on the $\phi^T$ --- the character
property (plus linearity) then gives its values in the whole 
of $\calH$. The beautiful discovery of Connes and Kreimer, who
introduced the above setting~\cite{Con.Kre:00}, is that $G$ is 
actually a (non-abelian) Lie group. Giving the group law
consists in giving the coordinates $\phi^{T_3}(g_3)$ 
of the product
$g_3= g_1 g_2$, as smooth functions of the
coordinates $\phi^{T_1}(g_1)$, $\phi^{T_2}(g_2)$ of the
factors. First, two simple examples,
\bae
\label{twoex}
\slts 
\phi^{\Too}(g_1 g_2)
\fe
\slts 
\phi^{\Too}(g_1) 
+
\phi^{\Too}(g_2) 
\ff
\slts 
\phi^{\Tto}(g_1 g_2)
\fe
\slts 
\phi^{\Tto}(g_1) 
+
\phi^{\Tto}(g_2)
+
\phi^{\Too}(g_1) \phi^{\Too}(g_2)
\, .
\nonumber
\eae 
Then, a not-so-simple example,
\bae
\label{nsse}
\phi^{\Ttht}(g_1 g_2)
\fe
\phi^{\Ttht}(g_1)
+
\phi^{\Ttht}(g_2)
+
2 \phi^{\Too}(g_1) \phi^{\Tto}(g_2)
\ff
 & &
{} +
\big( \phi^{\Too}(g_1) \big)^2 \phi^{\Too}(g_2)
\, .
\nonumber
\eae
Finally, the general formula,
\ble{genform}
\phi^T(g_1 g_2) = \phi^T(g_1)  + \phi^T(g_2) +
\sum_{\mbox{\small cuts } C}
\phi^{P^C(T)}(g_1) \phi^{R^C(T)}(g_2)
\, .
\ee
The sum is over all {\em admissible cuts} $C$ of the
tree $T$. An admissible cut can be either a {\em simple cut} or a 
{\em composite cut}. The former
involves cutting an edge of the tree and discarding the
half-edges produced (but {\em not} the vertices to which these are
attached). In this way, the tree is separated into two
subtrees, one that contains the root (denoted by $R^C(T)$
above), and one that ``falls to the floor''%
\footnote{%
A mental picture borrowed from D.{} Kreimer.%
} after the cut (denoted by
$P^C(T)$ above).
A composite cut, on the other hand,
involves cutting $k>1$ edges, with the
constraint that there is at most one cut on any path from the
root downwards. In this case $R^C(T)$ again denotes the
subtree containing the root but $P^C(T)$ now denotes the
collection $\{ T_i \}$, $1 \leq i \leq k$ of subtrees that fall to 
the floor, with  $\phi^{P^C(T)}$
denoting the product $\phi^{T_1} \phi^{T_2} \ldots \phi^{T_k}$.
It can be shown that the above group law is associative. 
It is easy to read off the coproduct of $\phi^T$
from~(\ref{genform}),   
\ble{genformcop}
\Delta(\phi^T) = \phi^T \ot 1 + 1 \ot \phi^T +
\sum_{\mbox{\small cuts } C}
\phi^{P^C(T)}  \ot \phi^{R^C(T)}
\, .
\ee
The units appearing in the \rhs{} denote the constant unit
function on $G$. 
From this expression one finds a recursive formula for the
antipode
\ble{recS}
S(\phi^T) = -\phi^T 
-
\sum_{\mbox{\small cuts } C}
S(\phi^{P^C(T)}) \phi^{R^C(T)}
\, .
\ee
The counit of all $\phi$'s is zero, while that of the unit
function is, of course, one. Given then any two 
characters (in particular, two renormalization schemes), one may 
form their product according to the above
formulas to obtain a new one. The counit just given implies that
the unit character, wrt{} this product, assigns the
value zero to all divergent integrals, while assigning 1 to
the unit function. This completes the Hopf algebra
structure of $\calH$, dual to the Lie group structure of $G$.

But, apart from aesthetics, what is it good for? The answer
lies, partly, in the following observation~\cite{Bro.Kre:99}.
Define a map $R$ that encodes the subtraction procedure, \eg,
by assigning to each $\phi^T$ the pole part
of its Laurent expansion, evaluated at $c=1$%
\footnote{%
In general, $R$ must satisfy
$R(xy)-R(R(x)y)-R(xR(y))+R(x)R(y)=0$ 
--- this guarantees that $S_R$ (see below) is multiplicative.%
}. Use $R$ to
define a {\em twisted antipode} $S_R$ via
\ble{twiS}
S_R(\phi^T) = -R(\phi^T)
-
R \left(
\sum_{\mbox{\small cuts } C}
S_R(\phi^{P^C(T)}) \phi^{R^C(T)}
\right)
\ee
(notice the similarity of the recursive structure with that
of~(\ref{recS})). Then all of renormalization's recursive
complexity is encoded neatly in the formula
\ble{renphiT}
\left. 
\phi^T_{\text{ren}}=S_R(\phi^T_{(1)})
\phi^T_{(2)}
\right|_{\epsilon=0}
\, .
\ee
\begin{example}{Renormalization of $\phi^{\Tto}$}{renTto}
From the coproduct
\ble{pTtocop}
\Delta(\phi^{\Tto})
=
\phi^{\Tto} \otimes 1 
+
1 \otimes \phi^{\Tto}
+
\phi^{\Too} \otimes \phi^{\Too}
\ee
and Eq.~(\ref{renphiT}), one gets
\ble{phiTtoren}
\phi^{\Tto}_{\text{ren}}
=
S_R(\phi^{\Tto}) 
+ \phi^{\Tto} 
+S_R(\phi^{\Too}) \phi^{\Too}
\, ,
\ee
the limit $\epsilon \rightarrow 0$ being implied. For the
twisted antipodes~(\ref{twiS}) gives
\bae
\label{twiStott}
S_R(\phi^{\Too})
\fe
-R(\phi^{\Too})
\ff
S_R(\phi^{\Tto})
\fe
-R(\phi^{\Tto})
-R(-R(\phi^{\Too}) \phi^{\Too})
\, .
\eae
Substituting these in~(\ref{phiTtoren}) gives
$\phi^{\Tto}_{\text{ren}}$ as a sum of 4 ($=2^2$) terms,
\ble{phiTtoren2}
\phi^{\Tto}_{\text{ren}}
=
\phi^{\Tto} -R(\phi^{\Tto}) 
-R(\phi^{\Too}) \phi^{\Too} 
+R(R(\phi^{\Too}) \phi^{\Too})
\, .
\ee
Evaluating the regularized integrals and expanding in Laurent
series in $\epsilon$ one finds (with some abuse of notation)
\bae
\label{Lsphioo}
\phi^{\Too}
\fe
\frac{1}{\epsilon} -a +
\big(
\frac{a^2}{2}+\frac{\pi^2}{6}
\big) \epsilon
+
\calO(\epsilon^2)
\\
\phi^{\Tto}
\fe
\frac{1}{2\epsilon^2}
-\frac{a}{\epsilon}
+a^2+\frac{5\pi^2}{12}
+\calO(\epsilon)
\label{Lsphito}
\, ,
\eae
so that, finally,
\ble{phirenf}
\phi^{\Tto}_{\text{ren}}
= 
\frac{a^2}{2}+\frac{\pi^2}{4}
\, .
\ee
\end{example}
\subsection{Primitive elements in $\calH$}
\label{peH}
As the last example probably makes clear, renormalizing multiply
nested integrals can be heavy work. A glance
at~(\ref{renphiT}) shows that the complexity of the task,
for a particular $\phi^T$,
depends on the number of terms in the coproduct of $\phi^T$.
In fact, one realizes that this dependence is even stronger by taking
into account~(\ref{twiS}), \ie, the fact that the complexity
of the twisted
antipodes produced by~(\ref{renphiT}) depends itself on that
of the coproduct. It would be nice then if the coproduct of
the $\phi$'s were simpler. But it isn't. The next best thing
is to look for a new set of coordinates on $G$ with simpler
coproducts. For example, one cannot ask for anything more from
$\phi^{\Too}$,
\ble{copToo}
\Delta(\phi^{\Too})
=
\phi^{\Too} \otimes 1 + 1 \otimes \phi^{\Too}
\, ,
\ee
(such a coproduct will be called {\em 1-primitive}) 
but $\phi^{\Tto}$ isn't as innocent, see~(\ref{pTtocop}) (and
Ex.~\ref{renTto} for the consequences). A
little experimentation though shows that
\ble{psitodef}
\psi^{\Tto} 
=
\phi^{\Tto} -\frac{1}{2} (\phi^{\Too})^2
\ee
also has 1-primitive coproduct (notice that the coproduct of a
product is the product of the coproducts). To renormalize 
$\psi^{\Tto}$ then,
two options are available. The first one is to write
\ble{badoption}
\psi^{\Tto}_{\text{ren}}
=
\phi^{\Tto}_{\text{ren}} -\frac{1}{2}
(\phi^{\Too}_{\text{ren}})^2
\, ,
\ee
and renormalize the $\phi$'s in the \rhs. Since
\ble{phiooren}
\phi^{\Too}_{\text{ren}}=
\left. 
\phi^{\Too}-R(\phi^{\Too})
\right|_{\epsilon=0}
= 
-a
\, ,
\ee
one finds, using the result of Ex.~\ref{renTto}
(Eq.~(\ref{phirenf}), $\psi^{\Tto}_{\text{ren}}=\pi^4/4$. The
second option is to use directly~(\ref{renphiT}), which holds
for functions of the coordinates as well since both the
coproduct and the twisted antipode are algebra homomorphisms 
--- this  gives
$\psi^{\Tto}_{\text{ren}}=\psi^{\Tto}-R(\psi^{\Tto})$. But a
Laurent expansion of $\psi^{\Tto}$ shows that
\ble{psiTtoL}
\psi^{\Tto}= \frac{\pi^2}{4} +\calO(\epsilon)
\, ,
\ee
\ie, $\psi^{\Tto}$ does not even have poles, so that
$R(\psi^{\Tto})=0$, and one recovers immediately the above
result. Clearly, there are good reasons to further investigate
the applicability of the second option. Consider, for example, 
renormalizing
a $\phi$ with ten vertices. The analogue of~(\ref{phiTtoren2})
would then contain $2^{10}=1024$ terms. On the other hand, 
imagine that appropriate additions to that $\phi$ turn it into a 
1-primitive $\psi$. Then only two terms would appear in its 
renormalization. Too good to be true? Read on. Some pertinent 
questions then are:
Can one find other
$\psi$'s with 1-primitive coproduct? Can all $\phi$'s be
traded for 1-primitive $\psi$'s? The answer is yes and no, in
that order. A little more experimentation, for example, shows that
\ble{psithree}
\psi^{\Ttho}
=
\phi^{\Ttho}
-\phi^{\Too} \phi^{\Tto}
+\frac{1}{3} (\phi^{\Too})^3
\ee
is also 1-primitive. On the other
hand, if a set of 1-primitive coordinates could be chosen on
$G$, then $G$ would be abelian, which it is not. The most one
can hope for then is to make an optimal choice of coordinates
that best adapts to the ``abelian directions'' on $G$.   
\subsection{Geometrization, part II}
\label{gpII}
``Ladder'' (linear) trees provide a convenient point of entry
to the problem of primitive elements in $\calH$.
It is easy to see that their coproduct is given by
\ble{laddcop}
\Delta(\phi^{(n)})=\sum_{k=0}^n \phi^{(k)} \otimes \phi^{(n-k)}
\, ,
\ee
where $\phi^{(r)}$ denotes the ladder tree with $r$ vertices
($\phi^{(0)}$ stands for the unit function). But this form of
the coproduct rings a bell. Consider the commutative algebra
of power series of the form $g= 1 +c_1 x + c_2 x^2 +
\ldots$,  with the usual product. Introducing
coordinates $\{ \xi_n \}$ on the space of such power series,
with $\ip{g}{\xi_n}=c_n$, one easily finds that dual to the
product of power series is the coproduct 
\ble{xicop}
\Delta(\xi_{n})=\sum_{k=0}^n \xi_{k} \otimes \xi_{n-k}
\ee
(compare with~(\ref{laddcop})). But series of the above form
can also be written as exponentials,
\ble{gexp}
g=e^{c'_1 x +c'_2 x^2 + \ldots}
\, .
\ee
Change now coordinates from $\xi$ to $\xi'$, such that
$\ip{g}{\xi'_n}=c'_n$. Under multiplication of series, the
constants $c'$ simply add up, implying the coproduct
$\Delta(\xi'_n)=\xi'_n \otimes 1 + 1 \otimes \xi'_n$, \ie, the
$\xi'_n$ are all 1-primitive. The obvious isomorphism with the 
Hopf algebra of ladder trees implies that the $\psi^{(n)}$
defined by
\ble{psiphiladd}
e^{\sum_{n=1}^\infty \psi^{(n)} x^n}= \sum_{r=0}^\infty
\phi^{(r)} x^r
\ee
are all primitive. (\ref{psiphiladd}) may be inverted to give
\ble{phipsiladd}
\psi^{(n)} = \frac{1}{n!} \frac{\partial^n}{\partial x^n} \log
\left( \sum_{m=0}^{\infty} \phi^{(m)} x^m \right) \bigg|_{x=0}
\, ,
\ee
putting the problem of the ladder 1-primitives, including the
ten-vertex one, to rest. 

This may look, so far, as a happy combinatorial accident. But
it actually points to the deeper geometrical origin of
1-primitiveness. There is no space here to delve into the
details but a sketch of some of the ideas will be given (for
details, see~\cite{Chr.Que.Ros.Ver:01}).
First, renormalization schemes are written as exponentials of
elements of $\mathfrak{g}$, the Lie algebra of $G$,
\ble{gexpZ}
g=e^{\sum_T c^T Z_T}
\, ,
\ee
where $Z_T$ are the generators of $\mathfrak{g}$. Then new 
coordinates $\psi^T$ are introduced, such that
\ble{psiTdef}
\ip{g}{\psi^T} = c^T
\, .
\ee 
These are, of course, the {\em normal coordinates} on $G$.  
Their coproduct is computed from the Baker-Cambell-Hausdorff
(BCH) formula,
\ble{psicop}
\Delta(\psi^A) =
\psi^A \ot 1 + 1 \ot \psi^A
+ \frac{1}{2} f_{B_1 B_2}^{\phantom{B_1 B_2}A} 
\psi^{B_1} \ot \psi^{B_2} + \dots
\, ,
\ee
where $f_{B_1 B_2}^{\phantom{B_1 B_2}A}$ are the structure
constants of $\mathfrak{g}$.
Some of the generators of $\mathfrak{g}$ cannot be written as
commutators. For the dual $\psi$'s the third term above, and
all higher, are zero, \ie, they are 1-primitive. This
observation points to a natural generalization of the concept
of 1-primitiveness. There are, for example, some other generators of
$\mathfrak{g}$ that {\em can} be written as commutators, but
not as double commutators. The dual $\psi$'s will have the
third term in the \rhs{} of~(\ref{psicop}) present but all
higher terms equal to zero --- they are to be called
accordingly 2-primitive. What enters naturally then in the
discussion of $k$-primitiveness is the lower central series of
$\mathfrak{g}$ which classifies its generators in classes
$\mathfrak{g}_k$, according to
the maximal number $k$ of nested commutators by which they can be
produced.   

But there is more to be gained from our geometric approach. 
For any function $f$ on $G$ one may
extend the exterior differential of $f$ at the identity,
$df|_e$, to a
left-invariant 1-form $\Pi_f$ on $G$. If $f$ is quadratic or
of higher degree in the $\phi$'s, $df|_e$, and hence $\Pi_f$,
vanishes. Given that the linear part of $\psi^T$ is $\phi^T$,
one concludes that $\Pi_{\psi^T}=\Pi_{\phi^T}$. For a primitive
$\psi^T$, on the other hand, the general formula 
\ble{PiLI}
\Pi_f=S(f_{(1)})df_{(2)}
\ee
shows that $\Pi_{\psi^T}=d\psi^T$, and hence,
$\Pi_{\phi^T}=d\psi^T$ 
is closed. Application of the inverse Poincar\'e lemma to
$\Pi_{\phi^T}=S(\phi^T_{(1)}) d\phi^T_{(2)}$ then provides an
expression of the 1-primitive $\psi^T$ as the co-cone
(potential) of the
closed $\Pi_{\phi^T}$. A little further trickery, that takes 
into account the fact that the coproduct of the $\phi$'s is 
linear in the $\phi$'s in the right tensor factor, results 
in an elegant formula,
\ble{psiPhiphi}
\psi^T = -\frac{1}{\Phi} S(\phi^T)
\, ,
\ee
where $\Phi$ counts the monomial order of the $ \phi$'s,
$\Phi(\phi^{T_1} \ldots \phi^{T_k})= k \, \phi^{T_1} \ldots
\phi^{T_k}$. 

Some {\em very} rough estimate of the savings in, say, CPU
time, from switching to the normal coordinates can be obtained
by assigning a cost of $2^k$ to a $k$-primitive element, 
regardless of the number of  vertices of its index, while
assuming that a
$\phi$ with $n$ vertices costs $2^n$. The
numbers $P_{n,k}$ of $k$-primitive elements with $n>k$ vertices
are given by the generating function~\cite{Bro.Kre:00}
\ble{Pnkgen}
P_k(x) \equiv \sum_{n=1}^\infty P_{n,k} x^n
= \sum_{s|k} \frac{\mu(s)}{k} \Bigl(
1-\prod_{n=1}^\infty \bigl( 1-x^{ns} \bigr)^{r_n}
\Bigr)^{k/s}
\, .
\ee
The sum in the \rhs \ above extends
over all divisors $s$ of $k$, including $1$ and $k$. $\mu(s)$
is
the M\"obius function, equal to zero, if $s$ is divisible
by a square, and to $(-1)^p$, if $s$ is the product of $p$
distinct primes $(\mu(1) \equiv 1)$.
The asymptotic behavior of $P_{n,k}$, for large values of
$n$ is~\cite{Bro.Kre:00}
\ble{asyPnk}
f_k \equiv \lim_{n \to \infty} \frac{P_{n,k}}{r_n} =
\frac{1}{c} \,
\Bigl(
1 - \frac{1}{c} \Bigr)^{k-1}
\, ,
\ee
where $c=2.95\dots$ is the Otter constant and $r_n$ is the
number of rooted trees with $n$ vertices. This is
good news, as the population of the CPU-intensive high-$k$
$\psi$'s is seen to be exponentially suppressed.
The ratio of
the total costs of renormalizing
all generators with $n$ vertices in the two bases then is
\ble{relco}
\rho_n = \frac{r_n 2^n}{\sum_{k=1}^{n-1} P_{n,k} 2^k}
\approx (c-2) \Bigl( \frac{c}{c-1} \Bigr)^{n-1}
\, ,
\ee
which soars to $5.4 \times 10^5$, a year's worth of
minutes, for $n=33$.

More insights, from a geometrical point of view, can be
found in~\cite{Chr.Que.Ros.Ver:01}. Very significant progress has
been made also with mostly algebraic means,
see~\cite{Foi:01} and, more 
recently,~\cite{Ebr.Guo.Kre:04,Ebr.Guo.Kre:04a}.
\section{Primitive Elements in the Hopf Algebra of Quantum Measures}
\label{GQM}
\subsection{Quantum mechanics as a quantum measure theory}
\label{qmaqmt}
A second example where geometry, in the guise of Hopf
algebras, illuminates problems of an algebraic/combinatoric
nature is furnished by Sorkin's proposal of a generalization
of quantum mechanics~\cite{Sor:94,Sor:97,Sor:97a} ---
Ref.~\cite{Chr.Dur:03} is followed closely in the sequel.
Consider the standard two-slit interference
experiment
and call $H$ the set of all electron histories (worldlines)
leaving
the electron gun and arriving at the detector at specified
time
instants (to avoid technicalities, consider $H$ to be
measurable).
Denote by $A$ ($B$) the subset of
$H$ consisting of all histories in which the electron passes
through slit $a$ ($b$), ignoring the possibility of the
electron winding around both slits. There are four
possible ways of blocking the two slits --- denote by
$P_{ab}$, $P_a$, $P_b$ and $P_0=0$  the corresponding
probabilities of arrival at the detector,
the last one corresponding to both slits being
blocked off. Sorkin's approach is to
consider these probabilities as the values of a certain
measure function $\mu$ defined on the set of subsets of $H$,
\eg,
$P_{a}=\mu(A)$. When mutually exclusive alternatives exist, as
when both slits are open, the union of the corresponding
(disjoint) subsets is to be taken, \eg, $P_{ab}=\mu(A \sqcup
B)$
($\sqcup$ denotes disjoint union). Physical
theories are distinguished by the measures they employ, for example,
classical mechanics uses a ``linear" measure $\mu_\text{cl}$,
satisfying the sum rule
\ble{sr2}
I^{\, \mu_\text{cl}}_2(A,B) \equiv \mu_\text{cl}(A \sqcup B)
-\mu_\text{cl}(A) - \mu_\text{cl}(B)=0
\, ,
\ee
and hence fails to predict any interference. Quantum
mechanics, on the other hand, uses $\mu_\text{q}$, satisfying
$I^{\, \mu_\text{q}}_2(A,B)\neq 0$. The
interesting observation by Sorkin is 
that in a three slit experiment (with eight
possibilities for blocking the slits), the probabilities
predicted by
quantum mechanics {\em do} satisfy the sum rule
\bae
\label{sr3}
I^{\, \mu_\text{q}}_3(A,B,C)
& \equiv &
\mu_\text{q}(A \sqcup B \sqcup C)
\ff
 & & 
{}-\mu_\text{q}(A \sqcup B)
- \mu_\text{q}(A \sqcup C)
- \mu_\text{q}(B \sqcup C)
\ff
 & &
{}+\mu_\text{q}(A)
+\mu_\text{q}(B)
+\mu_\text{q}(C)
\ff
 \fe 0
\, .
\eae
It is easy to show that
$\mu_\text{cl}$ also satisfies~(\ref{sr3}),
as a result of~(\ref{sr2}). There is an obvious generalization
to
the $k$-slit experiment, involving the symmetric functional
$I^\mu_k$, given by
\bae
\label{Ikdef}
I^\mu_k (A_1, \ldots,A_k)
& \equiv &
\mu(A_1 \sqcup \ldots \sqcup A_k)
\ff
 & &
{} -\sum_i
\mu(A_1 \sqcup \ldots \sqcup \hat{A_i} \sqcup \ldots \sqcup
A_k)
\ff
 & &
{}+\sum_{i<j}
\mu(A_1 \sqcup \ldots \sqcup \hat{A_i} \sqcup \ldots
\ff
& & 
\ldots \sqcup \hat{A_j} \sqcup \ldots \sqcup
A_k)
\ff
 & & \ldots
\ff
 & &
{} +(-1)^{k+1} \sum_i \mu(A_i)
\, ,
\eae
where symbols with hats are omitted and all $A_i$ are mutually
disjoint. Due to the recursion relation
\bae
\label{recIk}
I^\mu_{k+1}(A_0,A_1,\ldots,A_k)
\fe
I^\mu_k(A_0 \sqcup A_1,A_2,\ldots,A_k)
\ff
 & &
{}-I^\mu_k(A_0,A_2,\ldots,A_k)
\ff
 & & 
{} -I^\mu_k(A_1,A_2,\ldots,A_k)
\, ,
\eae
the sum rule $I^\mu_{k+1}=0$ follows from
$I^\mu_k=0$. It is natural now to contemplate a family of theories,
indexed by a positive integer $k$, defined by the sum rule
$I^\mu_{k+1}=0$, with $I^\mu_k \neq 0$
for the corresponding measure. In this scheme, classical mechanics 
is a $k=1$ theory while quantum mechanics corresponds to $k=2$.

The above formulas for $I^\mu_k$ need to be extended to the
general case, \ie, when the arguments are possibly overlapping
sets. For the $k=2$ case, Sorkin shows that bilinearity implies the
following equivalent forms
\bae
\label{k2over}
I^\mu_2
\fe
\mu(A \cup B) + \mu(A \cap B) - \mu(A \backslash B)
- \mu(B \backslash A)
\ff
 \fe
\mu(A \bigtriangleup B) + \mu(A) + \mu(B)
\ff
 & & 
{} -2 \mu(A \backslash
B)
-2 \mu(B \backslash A)
\, .
\eae
The symbol $\backslash$ above
denotes set-theoretic difference while $\bigtriangleup$
denotes symmetric difference.
\subsection{Coderivatives}
\label{Coder}
A brief excursion further into Hopf algebraic territory is necessary at
this point. 
One way of looking at the coproduct of a function is as
an indefinite translation. The 
right translation $R_g$ on the group is defined by $R_g(g')=g'g$. Its
pullback on functions $R_g^*(f) \equiv f_g$ is given by
$f_g(g')=f(g'g)=f_{(1)}(g') f_{(2)}(g)$. One infers that
$f_{(1)}(\cdot') f_{(2)}(g)$ is the right-translate of $f$ by
$g$,
while $f_{(1)}(\cdot') f_{(2)}(\cdot)$, a function of two
arguments, is the indefinitely translated $f$: its 
second
argument defines the translation while the first evaluates the
translated function (left translations can be similarly
handled
exchanging the two tensor factors of the coproduct). 
Introduce now the operator $\calL: \, \calA \mapsto
\calA \otimes \calA$,
defined by
\ble{Ldef}
\calL f = \Delta(f) - f \otimes 1
\, .
\ee
The above way of interpreting the coproduct shows that $\calL$
can be considered
an indefinite discrete derivative or {\em
coderivative} for short,
\bae
\label{idLd}
(\calL f)(g',\, g)
\fe
\ip{f_{(1)}\otimes f_{(2)} - f \otimes 1}{g' \otimes g}
\ff
 \fe
f(g'g)-f(g')
\, .
\eae
When $g$ is close to the identity, $g=e+X+\ldots$, with
$X$ in the Lie algebra of the group, $(\calL f)(\cdot',g)$ is
approximately (proportional to) the derivative of $f$ along the left
invariant vector field corresponding to $X$.
Higher order coderivatives $\calL^k f$ can similarly be
defined, with the
understanding that the successive applications of $\calL$ are
to be taken at the leftmost tensor factor,
\ble{Lkdef}
\calL^{k} f \equiv (\calL \otimes \id) \circ \calL^{k-1} f
\, ,
\qquad
k=2,3,\ldots
\, \, ,
\ee
so that, for example,
\bae
\label{L2ex}
\calL^2 f
& \equiv &
(\calL \otimes \id) \circ \calL f
\ff
 \fe
(\calL \otimes \id)  (f_{(1)} \otimes f_{(2)} - f \otimes 1)
\ff
 \fe
f_{(1)} \otimes f_{(2)} \otimes f_{(3)}
- f_{(1)} \otimes 1 \otimes f_{(2)}
\ff
 & &
 {}-f_{(1)} \otimes f_{(2)} \otimes 1
+ f \otimes 1 \otimes 1
\, .
\eae
Of particular interest is the evaluation of a
$k$-th order coderivative at the identity of the
group, $(\calL^k f)(e,\cdot, \ldots)\equiv (\calL^k f)(e)$,
\eg,
\bae
\label{L2ex2}
(\calL f)(e)
\fe
f - \epsilon(f) 1
\ff
(\calL^2 f)(e)
\fe
f_{(1)} \otimes f_{(2)} - f \otimes 1 -1 \otimes f 
\ff
 & &
{}+
\epsilon(f)
1 \otimes 1
\, .
\eae
It is now possible to introduce formally the
notion of {\em $k$-primitiveness}:
a function $f$ will be called {\em $k$-primitive} if all its
coderivatives of order $r$ at the identity, $(\calL^r f)(e)$, 
with $r > k$ are
equal to zero, while $(\calL^k f)(e)$ is not.
\subsection{Quantum measures and
$k$-primitiveness}
\label{qmkp}
\subsubsection{The abelian group of histories}
\label{tagh}
Referring back to the $k$-slit experiment, call $H$ 
the set of histories available to a particle,
taken as a measurable set for simplicity.
One may deal with a given subset $A$ of $H$ in terms of its
{\em characteristic function}, defined by $\chi_A(x)=1$
if
$x\in A$, $\chi_A(x)=0$ if $x \in H \backslash A$.
Denote by $G$ the set of all 
linear combinations of
characteristic functions of measurable subsets of $H$%
\footnote{%
These are known as {\em simple} functions on $H$.}.
A typical element $g$ of $G$ is of the form
$g= \lambda_1 \chi_{A_1} + \lambda_2 \chi_{A_2} + \ldots$,
where the
$A_i$ are measurable subsets of $H$ and $\lambda_i \in
\mathbb{C}$.
$G$ may be turned into an abelian group with the group law given
by addition. Then the identity $e$
is the zero function, $e= \chi_\emptyset=0$ and the inverse of 
$g$ is $-g$.

As in Sorkin's approach, a physical theory derives its
probabilities from a measure function $\mu$, defined now on
$G$, \eg, $P_a=\mu(\chi_A)$ in the two-slit experiment. In the
presence of
mutually exclusive alternatives, the {\em sum of the
characteristic functions of the corresponding subsets} is to
be taken --- this corresponds to disjoint union in terms of the 
subsets themselves, as in~\cite{Sor:94},
\cite{Sal:02}. What is attractive though in working with
characteristic functions is that by extending
this definition (\ie, addition of the characteristic
functions) to
non-disjoint subsets we recover the rather
complicated interference term~(\ref{k2over}) and its
generalizations. Consider for example a quadratic
functional $\mu^2$, with $\mu$ additive, evaluated on two
overlaping
subsets $A$ and $B$  --- the resulting interference term is
\bae
\label{recint}
I^{\mu^2}_2
\fe
\mu(\chi_A + \chi_B)^2
-
\mu(\chi_A)^2 - \mu(\chi_B)^2
\ff
 \fe
 2 \mu(\chi_A) \mu(\chi_B)
\ff
 \fe
2 \big(
\mu(\chi_{A \backslash B}) \mu(\chi_{B \backslash A})
+ \mu(\chi_{A \backslash B}) \mu(\chi_{A \cap B})
\ff
 & &
{}+ \mu(\chi_{A \cap B}) \mu(\chi_{B \backslash A})
+ \mu(\chi_{A \cap B})^2
\big)
\, ,
\eae
where, in the last step, the substitution $\chi_A=\chi_{A \backslash
B} +
\chi_{A \cap B}$ was used, and similarly for $\chi_B$.
On the other hand, the first, for example, of~(\ref{k2over})
becomes
\ble{calcint}
I^{\mu^2}_2
=
\mu(\chi_{A \cup B})^2 + \mu(\chi_{A \cap B})^2 -\mu(\chi_{A
\backslash B})^2 - \mu(\chi_{B \backslash A})^2
\, .
\ee
Substituting
$\chi_{A \cup B}=\chi_{A \backslash B} + \chi_{B \backslash A}
+
\chi_{A \cap B}$ and expanding one recovers the right hand
side
of~(\ref{recint}).

The quantum measures considered up to now are elements of the
commutative Hopf algebra 
$\calA \equiv C^\infty(G)$ of smooth functions on $G$.
Their counit vanishes, since $\mu(\emptyset)=0$.
The linearity of the classical measure, Eq.~(\ref{sr2}), 
implies
that $\mu_\text{cl}(\chi_A + \chi_B)= \mu_\text{cl}(\chi_A) +
\mu_\text{cl}(\chi_B)$, or, in terms of the coderivative,
\ble{muclH}
0 = (\calL^2 \mu)(e) = \mu_{\text{cl} \, (1)}
\otimes \mu_{\text{cl} \, (2)}
- \mu_\text{cl} \otimes 1 - 1 \otimes \mu_\text{cl} +
\epsilon(\mu_\text{cl}) 1 \otimes 1
\, ,
\ee
the last term being zero. Hence,
$\mu_\text{cl}$ is a 1-primitive element of $\calA$.
More generally, the following lemma holds
\begin{lemma}
The symmetric functionals $I^\mu_k$, defined in
Eq.~(\ref{Ikdef}),
coincide with the $k$-th order coderivatives $(\calL^k
\mu)(e)$
of Eq.~(\ref{Lkdef}).
\end{lemma}
The straightforward inductive proof is left as an exercise. 
The main result of this section may now be stated
\begin{prop}
In the algebra $\calA$ of functions on $G$,
every $k$-primitive element
is a $k$-th degree polynomial in 1-primitive elements.
\end{prop}
The proof can be given in a number of ways. For example, it is
easily established that 
$\calA$ is
a cocommutative graded connected Hopf algebra. This means that
the coproduct is symmetric under exchange of the two tensor
factors, that there is a grading respected by the coproduct
($k$-primitiveness) and that the only elements with zero grade
are numbers. Then, application of the Milnor-Moore 
theorem~\cite{Mil.Moo:65}, gives that $\calA$ is
isomorphic to the universal enveloping algebra of its
subalgebra
of 1-primitive elements, which establishes the proposition. 
Alternatively, and closer to the spirit of this paper, one may
introduce normal coordinates on $H$. Because $H$ is abelian
the normal coordinates are 1-primitive. Then the vanishing
at the identity of all coderivatives, of order higher than $k$ 
(but not that of order $k$), of a function 
of the coordinates implies that the function is
a polynomial of order $k$ in the coordinates, q.e.d. 
\section{Conclusions}
\label{Conc}
It has been argued that a geometric setting, cast in Hopf
algebraic language, might be efficient in dealing with
problems of a combinatoric nature. Two examples were used to 
illustrate the
point: 
\begin{itemize}
\item
A toy-model renormalization is simplified significantly
by  introducing normal
coordinates on the infinite dimensional Lie group of 
renormalization schemes of Connes and Kreimer, while 1-primitive
functions on the group are shown to
correspond to closed  left-invariant 1-forms.
\item
Quantum measures of $k$-th order, proposed by Sorkin as
generalizations of quantum mechanics, are described as
$k$-primitive functions on the group of histories associated
to a particular experiment. As a result, they are shown to be
polynomials of order $k$ in additive (\ie, classical-like)
measures.  
\end{itemize} 
It would be nice to see this list grow longer fast. 
\section*{Acknowledgements}
\label{Ackn}
I would like to thank the organizers of the V Workshop of the
Division of Gravitation and Mathematical Physics of the
Mexican Physical Society in Morelia, Mexico, for their invitation. 
Warm thanks are
also due to Denjoe O'Connor, for hospitality and financial
support while at
DIAS, Ireland, where the present work was written. Partial
support from DGAPA-PAPIIT grant IN 114302 and CONACyT grant
4M1208-F is also acknowledged.  
\longpage

\end{document}